**Extreme ultraviolet lithography reaches 5 nm resolution**

*I. Giannopoulos, I. Mochi, M. Vockenhuber, Y. Ekinci & D. Kazazis*

*Paul Scherrer Institute, 5232 Villigen-PSI, Switzerland*

## Abstract

Extreme ultraviolet (EUV) lithography is the leading lithography technique in CMOS mass production, moving towards the sub-10 nm half-pitch (HP) regime with the ongoing development of the next generation high-numerical aperture (high-NA) EUV scanners. Hitherto, EUV interference lithography (EUV-IL) utilizing transmission gratings has been a powerful patterning tool for the early development of EUV resists and related processes, playing a key role in exploring and pushing the boundaries of photon-based lithography. However, achieving pattering with HPs well below 10 nm using this method presents significant challenges. In response, our study introduces a novel EUV-IL setup that employs mirror-based technology and circumvents the limitations of diffraction efficiency towards the diffraction limit that is inherent in conventional grating-based approaches. We present line/space patterning of HSQ resist down to HP 5 nm using the standard EUV wavelength 13.5 nm, and the compatibility of the tool with shorter wavelengths beyond EUV. The mirror-based interference lithography tool paves the way towards the ultimate photon-based resolution at EUV wavelengths and beyond. This advancement is vital for scientific and industrial research, addressing the increasingly challenging needs of nanoscience and technology and future technology nodes of CMOS manufacturing in the few-nanometer HP regime.

Over the past four decades there have been unprecedented advancements in the production of integrated chips with increasing computational power. This became possible by the scaling of the semiconductor devices, as dictated by G. E. Moore in 1965[1]. Following his predictions, the semiconductor industry has continued the miniaturization of the transistor and the increase of the device density per integrated chip. This would not have been technologically possible and economically viable without a patterning technique capable of printing a large number of devices on a chip in a parallel manner, namely, top-down photolithography. There is a large variety of lithography techniques with very high resolution such as electron[2] or focused ion beam lithography[3], nanoimprint lithography[4,5], techniques based on scanning probes[6-9] or scanning tunneling microscopes[10]. Although techniques like scanning tunneling lithography have demonstrated atomic scale patterning[11], it is impractical to integrate them in high volume manufacturing (HVM), because they are serial techniques and therefore incompatible with industrial throughput[12]. Photolithography is a parallel technique where the image of a specific mask is projected on a thin film of photosensitive material resulting in a local solubility change. The soluble part of the photosensitive material (photoresist) is washed away in a special developer solution. The mask pattern transferred to the photoresist can be further transferred to the silicon substrate creating functional electronic components.

Apart from being a key technology for the production of semiconductor electronics in HVM, photolithography has boosted the miniaturization and the performance of integrated semiconductor devices, hence the evolution of computational hardware towards our modern standards. State-of-the-art integrated circuit production has relied on photolithography tools, steppers or scanners for nearly four decades.

The resolution (R) of any imaging system is given by the Rayleigh criterion

$$R = \frac{k_1 \lambda}{\text{NA}}, \tag{1}$$

where NA is the numerical aperture of the imaging lens system, $\lambda$ is the wavelength of the light, and $k_1$ is a process-dependent parameter linked to the illumination system, the mask architecture and other variables[13]. Over the years, feature sizes have been reduced by first increasing the NA of the imaging lens system, then decreasing the parameter $k_1$ and eventually decreasing the wavelength of the employed light source. The wavelength had reached a prolonged plateau at 193 nm (ArF excimer laser), because the attempts for further reduction to 157 nm ($F_2$) were abandoned[14] due to the birefringence observed in $CaF_2$, the only practical replacement of quartz as a lens for this wavelength[15]. Meanwhile, numerous solutions have been introduced to increase the feature density and the patterning resolution, such as optical proximity correction[16], phase-shift masks[17] and the highly effective yet cost-inefficient multiple patterning[18]. A pivotal development was immersion lithography, where a high refractive index liquid was introduced in the air gap between the final lens and the wafer to increase the NA to values greater than 1 and boost the resolution[19,20].

However, the computational challenges of recent years have led to a surging demand for power efficient and high-performance semiconductor devices and spurred global research in advancing high-resolution lithography. This was achieved by reducing the wavelength to 13.5 nm that falls in the extreme ultraviolet (EUV) range. EUV lithography entered HVM in 2019 leading to a further plummet of the minimum feature size. The more than tenfold reduction in wavelength was achieved through extensive academic and industrial research and development efforts. These focused on creating innovative EUV sources with sufficient power and redesigning the illumination and projection optics of EUV scanners to operate in reflective mode. This was necessary because EUV photons are absorbed by most materials, including air, making the use of refractive optics practically impossible. In a modern EUV scanner, the projection takes place through a cascade of multilayer-coated mirrors in a low pressure hydrogen environment[21], to protect the optics from contamination[22,23]. The NA of the systems in production is 0.33, leaving substantial room for improvement on that aspect. Systems with an NA of 0.55 (high-NA systems) are currently under development, while the possibility of even higher NAs of 0.75 or 0.85 (hyper NA) is also explored[24]. High-NA ultimately aims for patterning at half pitch (HP) resolutions down to 8 nm by 2028[25]. Despite optimization, absorptions from reflective optics dramatically reduce the photon flux on the wafer, raising the need for high-power sources and high-sensitivity resists. Although there are projections for EUV source powers up to 800 W[26], it is still of paramount importance to develop EUV photoresist materials with high sensitivity, while maintaining ultra-high resolution and low line width roughness (LWR). In this demanding ecosystem, EUV interference lithography (EUV-IL) has been instrumental in the development of EUV photoresists and processes[27-32]. This method provides a high-resolution aerial image in a cost-effective manner, for the timely development of photoresist materials and processes even before the availability of EUV scanners.

In this work, we extend the capabilities of EUV-IL by introducing an alternative mirror-based method. In EUV mirror interference lithography (MIL), two mutually coherent beams are reflected by two identical mirrors. The reflected beams create an interference pattern with a pitch that depends on the grazing angles and the wavelength. Due to the absence of any diffractive elements and the high reflectivity of the Ru mirrors that we use, MIL exposures are characterized by very high efficiencies and are, therefore, less prone to thermomechanical drifts. Consequently, the MIL method is capable of sub-10 nm HP resolutions. In addition, it provides a means to expose a photoresist with variable contrast in a well-controlled manner and in a single exposure. Our results demonstrate that it is possible to reach ultimate resolutions with photon-based lithography and open new avenues for research in the field of EUV lithography and photoresist materials. The following paragraphs offer an exploration of the key aspects of MIL methodology, starting with the description of the experimental setup and the theoretical background, followed by an in-depth analysis of the variable contrast across the imaging field. Moving on to the experimental results, we present sub-10 nm HP lithographic exposures down to 5 nm HP with 13.5 nm light and a case-study analysis of the image contrast. Finally, we showcase the compatibility of MIL with shorter wavelengths beyond EUV and discuss the potential of the technique.

## Results

### EUV Mirror Interference Lithography

EUV-IL is a lithography technique based on the interference of two or more mutually coherent beams which can create periodic aerial images, such as line/space patterns and contact holes or pillars as well as more complex structures such as kagome[33] or penrose[34] patterns. Unlike commercial scanners, EUV-IL tools do not employ complex optics. Instead, they use transmission diffraction grating masks (see supplementary information **A** & Fig. S1). To minimize EUV absorption, the masks are made of thin silicon nitride membranes (80 nm) that absorb roughly 50% of the incoming light[35]. To obtain a line/space pattern, two identical diffraction gratings with a predefined distance from each other are fabricated on the membrane. When the gratings are irradiated with EUV synchrotron light, the diffracted beams interfere creating a periodic aerial image with a pitch corresponding to a fraction of the one on the mask gratings. When a wafer is positioned at a specific distance from the mask, where the diffracted beams overlap, the interference pattern is recorded in the photoresist. This pattern covers a relatively large area (on the order of 100x100 $\mu m^2$) with a substantial throughput, boasting high and pitch independent contrast and absence of depth of focus limitations. Synchrotron light stands out as an ideal source for EUV-IL due to its high photon-flux and coherence.

The EUV-IL tool at the XIL-II beamline of the Paul Scherrer Institute has shown line/space half-pitch resolutions down to 6 nm[36]. Owing to its high resolution, absence of complex optics, easy access, low outgassing and no material restrictions for exploratory resist systems, EUV-IL has been extensively used in the development and evaluation of photoresist materials for EUV lithography, even before commercial tools became available[30,37,38]. Additionally, EUV-IL has been particularly attractive for a plethora of scientific applications that require periodic nanopatterning over relatively large areas[39-48].

Despite the high brilliance of the synchrotron source, diffraction gratings suffer from limited diffraction efficiency. Typically, these gratings are fabricated by patterning a resist layer directly into line/space arrays without any further pattern transfer. Hydrogen silsesquioxane (HSQ) gratings present a favorable choice due to their reasonable diffraction efficiency, stability under EUV irradiation and high-quality patterning with electron beam lithography (EBL). So far, the efforts to optimize materials and increase the diffraction efficiency, such us the use of bilayer stacks of spin-on carbon and HSQ[49], were met with limited success and finding a solution to the dramatic reduction of diffraction efficiency for patterning below HP 10 nm is still an object of research. Low diffraction efficiency implies higher exposure times that, in turn, make the exposure more prone to thermal drifts and mechanical vibrations, leading to increased aerial image blur, hence limited resolution. Besides diffraction efficiency limitations, the nanofabrication of high-quality gratings on thin silicon nitride membranes becomes increasingly challenging for sub-10 nm HP resolution, due to EBL resolution limitations and pattern collapse.

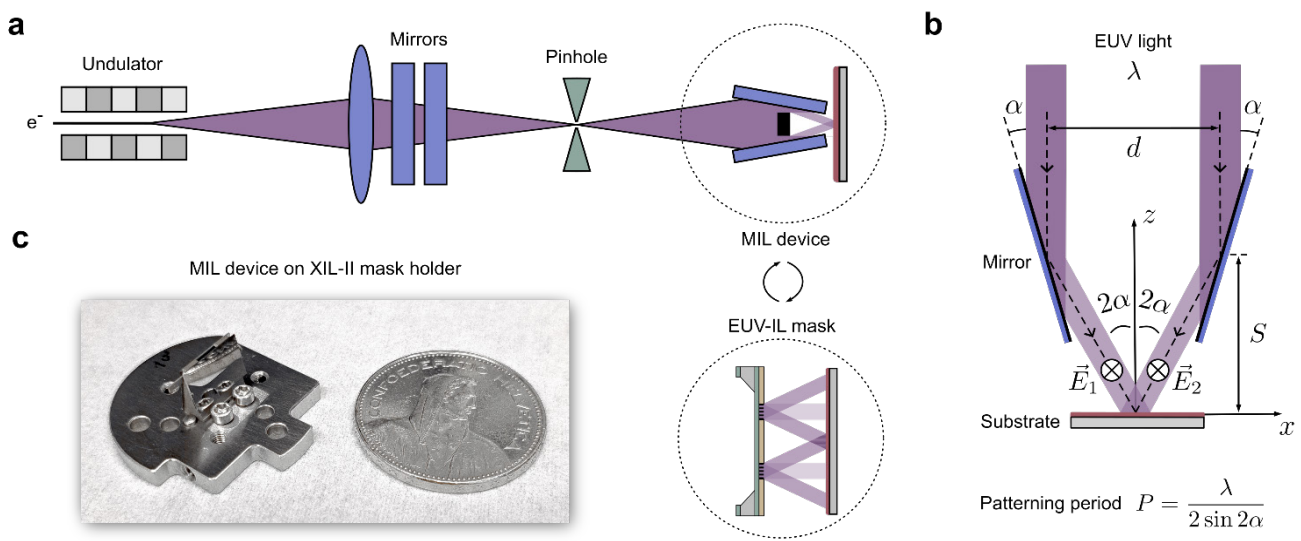

**Figure 1 | Mirror interference lithography.** (**a**) Schematic representation of the XIL-II beamline at the Paul Scherrer Institute. EUV light with tunable wavelength is generated by an undulator and gets reflected off a series of mirrors for high harmonics suppression and focusing. The beam is focused on a pinhole (spatial filter) and the spatially coherent beam subsequently illuminates the imaging module (mask with gratings or MIL device). (**b**) MIL device along with the principle of operation. After passing through the pinhole, the EUV beam is split in two parts by a physical photon blocker. After reflecting on mirrors inclined at a specified grazing angle, the two beams interfere on the resist substrate. (**c**) Photograph of a MIL device mounted on an EUV-IL mask holder, next to a 5 Swiss francs coin.

To avoid these challenges and in pursuit of the ultimate resolution for photon-based lithography, we shift away from the transmission gratings approach and introduce EUV mirror interference lithography. Figure 1 shows a schematic design of the XIL-II beamline that we developed[50]. An electron beam sourced by the Swiss light source (SLS) accelerator is guided through an undulator that generates a highly coherent and brilliant EUV light beam. A set of reflective optics filter, shape and focus the beam on a pinhole. The spatially coherent beam propagates towards the endstation, where the MIL device and the substrate are positioned. Within the MIL device, a mechanical photon blocker divides the beam and two mirrors located at a distance $d$ from each other reflect the two coherent beams that have a width $w$. The mirrors are positioned at a grazing angle $\alpha$ with respect to the incident beam. The general expression[51] of the interference pattern intensity as a function of the position $x$ away from the centerline for a given mirror angle $\alpha$ and wavelength $\lambda$ is given by equation (2), where $A$ is the amplitude of the electric field vector. We assume plane waves with transverse electric polarization (TE), meaning that the component of the electric field is perpendicular to the plane of incidence as defined by the propagation vector and the surface normal (see Fig. 1**b**). The reflected beams overlap at a distance $S = d/(2 \tan 2\alpha)$ from the center of the reflection area and form an interference pattern with a pitch $P$ given by equation (3). The derivation of these equations can be found in the supplementary information **B**. The exact grazing angles can vary uniquely for each MIL device, because of micromachining inaccuracies and the manual attachment of the mirrors. Nevertheless, the technique exhibits a high tolerance to such geometrical nonidealities, ensuring lithography results even in the presence of deviations from the intended pitch value (see supplementary information **C**).

$$I(x) = A^2 \left| e^{i\frac{2\pi x}{\lambda}sin(2\alpha)} + e^{i\frac{2\pi x}{\lambda}sin(-2\alpha)} \right|^2 \qquad (2)$$

$$P = \frac{\lambda}{2 \sin 2\alpha} \qquad (3)$$

Each part of the beam in this device undergoes only one reflection on a highly reflective planar mirror. In contrast to grating-based EUV-IL, there is neither a diffraction process with limited efficiency nor any absorbing membrane involved. This leads to considerably shorter exposure times in comparison to the grating-based method. The device is fully compatible with our standard EUV-IL system, requiring no modifications to the endstation and the beamline infrastructure. In addition, MIL devices are much more durable, as opposed to the gratings on membranes that are fragile and prone to degradation due to beam damage and contamination. This is due to their robust metallic structure, with the mirrors being the only components prone to degradation, yet easily replaceable with Ru-coated Si chips (see Methods).

Nevertheless, the fabrication of the MIL setup is not simple and requires a particularly rigorous micromachining process with adequate accuracy and precision according to the design. Excessive misalignments can easily render the setup completely unusable. Moreover, its positioning with respect to the beam is subject to relatively narrow misalignment tolerances. In terms of throughput, we note that while our typical EUV-IL mask features gratings with 5 or 6 different pitches in one exposure region, the fixed grazing angle limits MIL to only one pitch per device. However, MIL targets applications where ultimate resolution is required and is a very important asset in the portfolio of interference techniques available at our endstation. The most prominent difference between MIL and different EUV-IL methods is the contrast variation across the imaging area, a topic new and exclusively present to the MIL technique. We will study this phenomenon in depth and demonstrate the substantial value that it adds to the current capabilities of EUV-IL.

**Contrast range and variable NILS**

The intensity contrast of the interference pattern, as for any aerial image, is expressed as the ratio $(I_{max} - I_{min})/(I_{max} + I_{min})$. An ideally monochromatic beam would give maximum contrast across the complete area of the overlapped beams, as shown in equation (4). However, the spectral content of our illumination system causes an intensity modulation in the interference pattern across the field of view. This leads to a gradual contrast reduction as a function of the distance from the center with maximum contrast at the area of the centerline (see Fig. 2**a**). The effect of the bandwidth on the contrast reduction at positions away from the geometrical center of the beam overlap can be derived and numerically calculated for different wavelengths and grazing angles. Specifically, equation (2) gives (see supplementary information **B**):

$$I(x) = 4A^2 \cos^2\left(\frac{2\pi}{\lambda}\sin(2a)\,x\right).$$

(4)

Normalized image log-slope (NILS) is the standard metric that characterizes the edge definition of a pattern[52]. It is the slope of the aerial image intensity at the border of the pattern area, normalized by the intensity and the nominal linewidth.

$$\text{NILS} = \frac{1}{I}\frac{dI}{dx}w.$$

(5)

The general expression using the MIL intensity function, equation (6), is: (see supplementary information **D**)

$$\text{NILS} = -\pi \, tan\left(\frac{\pi}{p}x\right). \tag{6}$$

NILS is locally calculated at the line edges with half-pitch linewidth, meaning every ¼ and ¾ of the pitch. Solving equation (6) at the sequence of positions $x = (2n-1)\frac{P}{4}$, where $n$ is a nonzero integer, yields $|(-1)^n \pi| = \pi$. This constant and pitch-independent NILS number is a distinctive feature of interference lithography, and it also applies to the conventional transmission gratings case.

There is, however, a fundamental difference between the grating-based and the mirror-based IL when dealing with non-monochromatic light. In grating-based EUV-IL, the diffraction angle depends on the wavelength, so does the interference pattern, resulting in a wavelength-independent pitch (see supplementary information **A**). Oppositely, the interference pitch and the wavelength are proportional in MIL as shown in equation (2). Differences in the optical path lengths of the temporally incoherent frequency components arise along the finite width of the interference area. Our EUV beam is quasi-monochromatic, because the full width at half-maximum $\Delta f$ is much smaller than the central frequency $f_0$ that corresponds to the 13.5 nm wavelength. The power spectral density has a Gaussian distribution around $f_0$ with a bandwidth $\Delta f/f_0 = 4\%$. The normalized Gaussian expression of equation (7) describes the relative irradiance contributions of the involved optical frequencies. The total intensity is calculated as the integral over all the spectral components, which is approximated by the sum of the discrete frequencies given adequately fine slicing $\delta f$ in equation (8).

$$S(f) = \frac{1}{b\sqrt{\pi}} \, e^{-\left(\frac{f-f_0}{b}\right)^2}, \;\; \text{with} \;\; b = \frac{\Delta f}{2\sqrt{\ln 2}} \tag{7}$$

$$I(x) = \int_{-\infty}^{+\infty} I(x)S(f)df \cong \sum_{n=1}^{N} I(x)S(f)\delta f \tag{8}$$

Figure 2**a** shows the calculated intensity versus lateral position for a MIL device that gives 8.7 nm HP lines. According to the definition of NILS, one can calculate the slope and the value of the intensity curve at the linewidth borders for each period. Figure 2**b** shows the NILS number that corresponds to each interference line within the field of view for 3 different bandwidths of 1%, 2%, and 4%. The NILS of $\pi$, that corresponds to monochromatic light, is met only at the centerline, followed by a gradual decrease for every subsequent intensity peak. Even though the positions of these points change with the grazing angle, the NILS of each line remains independent of the pitch, meaning that the calculated curves shown in Fig. 2**b** are the same for all MIL devices, depending only on the bandwidth.

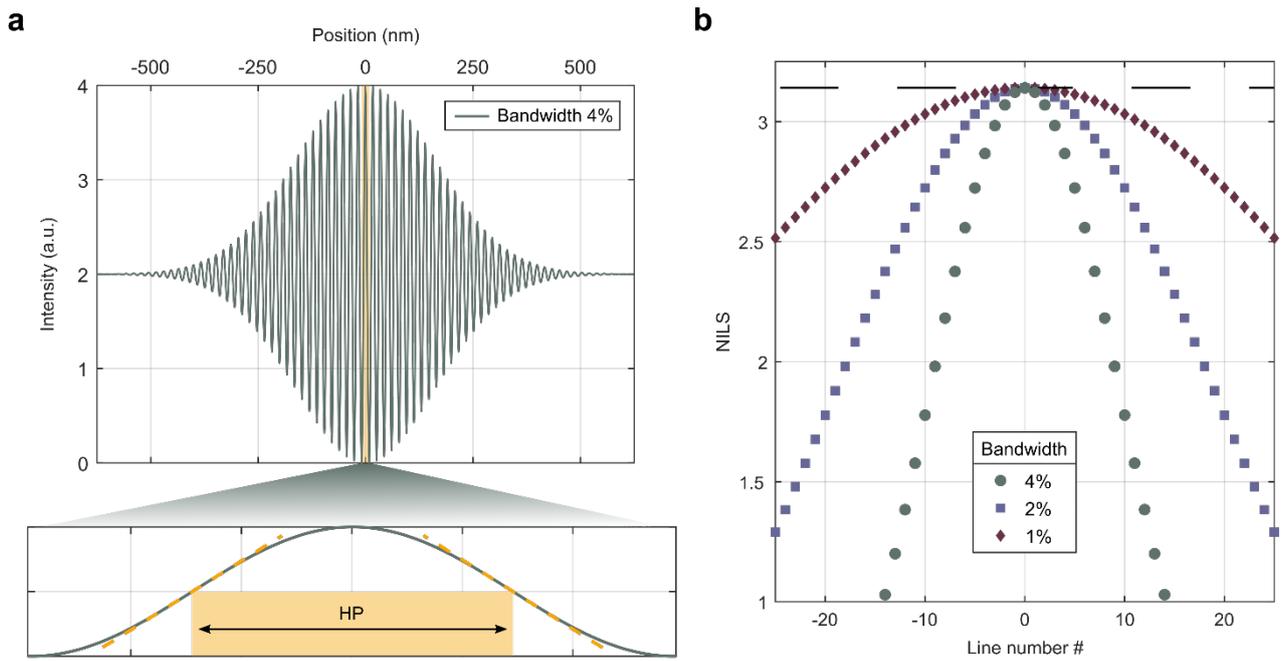

**Figure 2 | The characteristics of a MIL aerial image.** (**a**) Numerical simulation for an EUV beam with 4% bandwidth: The amplitude of the Interference pattern as a function of distance from the centerline for a HP 8.7 nm MIL device. NILS is calculated at each period with slope fits at the half-pitch borders. (**b**) Pitch-independent NILS number for each line away from the central one that has the maximum value $\pi$. The spectral content of the beamline used in this work has a Gaussian distribution with 4% FWHM and is plotted in comparison with the 1% and 2% ones. A gentler NILS decline can be obtained with smaller bandwidth values as well as a broader patterned area and a finer spacing between the available NILS levels.

Although the contrast loss may seem like an undesirable effect, it is, in fact, of great importance for research and development. Future reflective optics will push the lithographic resolution limit lower, but at the expense of having extremely shallow depth of focus and NILS numbers below 3 for HP lower than 15 nm[23,24]. Consequently, there is an urgent requirement for improved focus capabilities and systems that preserve wafer flatness throughout the process. In that context, being able to map the effect of contrast loss on a resist with a single exposure is of great relevance, because NILS numbers reduce dramatically at smaller pitches even at the best focus conditions[24]. A scanner performs exposures at different NILS conditions, but the exact NILS depends on many factors such as the illumination system, the focus, and the mask, therefore, it cannot be easily controlled. There have been previous attempts to tune the NILS in EUV-IL, so it matches the one of a scanner, by adding background (flare) to reduce the contrast, but this required multiple exposures[53]. Oppositely, a single MIL exposure contains multiple NILS numbers, hence, one can selectively characterize the lines that correspond to the contrast conditions of a given process in a simple manner.

**Experimental results**

Figure 3 shows the calculated reflectivity of a 10 nm Ru mirror as a function of grazing incidence angle[35], for the photon energy of 91.9 eV (13.5 nm wavelength). A 3 nm native silicon oxide layer and the silicon substrate are included in the calculation as well as the surface roughness. Based on these values, one can calculate the expected light intensity that arrives at the wafer as follows. If the incoming light intensity is measured $I_i$ ($W/cm^2$) before the device e.g., with a photodiode, then the intensity at the overlapping area on the substrate (see Fig. 1) is $2I_iR_a$, where $R_a$ is the mirror reflectivity. The angle-dependent $R_a$ factors, denoted by dots in Fig. 3, correspond to existing MIL devices used in this work. But this product describes only the superposition of the incoherent interfering waves. The spectral content of our beam confines the constructive interference in the limited area shown in Fig. 2**a**, where there is a fourfold increase in intensity according to equation (4). In equation (9) we define the ratio between the incoming intensity $I_i$ and the intensity on the resist $I_r$ at the patterning area as the "tool factor" $\mathrm{TF} = 4R_a$. This value directly measures the efficiency of IL, as it effectively shows the fraction of incoming photons per unit area utilized for patterning the photoresist. Tool factors can be measured experimentally by exposing a specific and stable resist with previously measured sensitivity, allowing for the estimation of the effective mirror reflectivity. Various factors such as surface roughness, oxidation, contamination, and thickness homogeneity may lead to deviations from the simulated levels numbers.

$$I_r = 4I_iR_a = I_i\mathrm{TF} \tag{9}$$

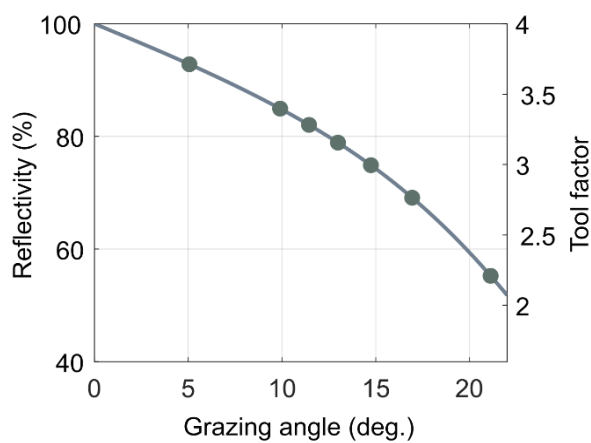

**Figure 3 | Mirror reflectivity.** Calculated reflectivity as a function of grazing angle for a low-roughness, 10 nm Ru film and a photon energy that corresponds to EUV light with wavelength of 13.5 nm. The markers show the grazing angles of the MIL devices used in this study.

To evaluate the performance and capabilities of MIL, we use HSQ, a commercial high-resolution resist that has been extensively used in EBL[54], but can also be exposed with EUV light[28]. HSQ falls short of the industrial sensitivity requirements for EUV resists by an order of magnitude, making the exposure times too long for industrial integration. Nevertheless, we use it as a benchmarking resist due to its ultra-high resolution below 10 nm HP, better than what most state-of-the-art EUV photoresists can achieve. The low sensitivity of HSQ is not a problem for MIL owing to its high efficiency that keeps the exposure time at only a few seconds.

Figure 4 shows scanning electron microscopy (SEM) images of HSQ lines with sub-10 nm HP patterned using MIL. We highlight the HP 6 and HP 5 images produced by MIL devices with 17° and 21.2° grazing angles, respectively. This resolution, utilizing the industrial standard EUV wavelength, establishes the new record in photon-based nanolithography.

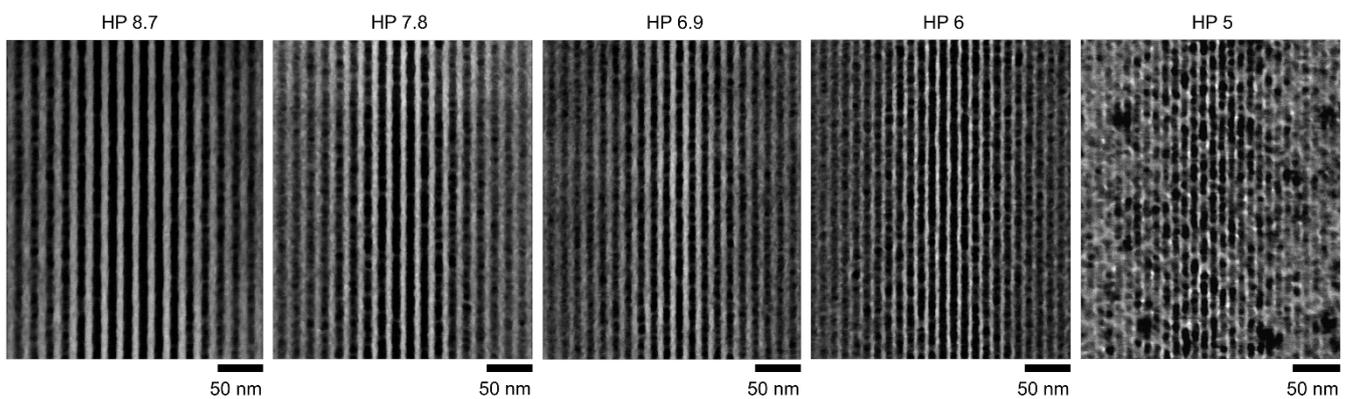

**Figure 4 | MIL exposures using the industry standard EUV wavelength of 13.5 nm.** SEM images of exposed HSQ photoresist using the MIL devices that are designed for sub-10 nm HP patterning. The HP that is linked to the grazing angle $\alpha$ (see Fig. 1) is unique for each MIL device at the 13.5 nm wavelength. The imaging area is centered at the maximum-contrast centerline (NILS = $\pi$) and shows the effect of contrast loss away from the centerline.

It is important to note that the reported SEM images of Fig. 4 were obtained immediately after a standard HSQ wet development process (see Methods), without any scum mitigation steps during or after the process. Undoubtedly, there is room for improvement in the development of the resist. For instance, more effective surfactants, beyond those already incorporated in the developer solution, could further reduce water tension, and allow for better access of the solution into such confined spaces. Additionally, optimizing the processing of HSQ, including potential adjustments to development time and temperature, could improve the LWR and reduce any residual scum. However, it is beyond the scope of the current study to explore such optimizations related to a specific photoresist; our primary aim is to conclusively highlight the formation of a high-resolution aerial image based on mirror interference of EUV light and, consequently, the ultimate patterning capabilities of the technique.

In Fig. 5 we demonstrate the analysis of a sample SEM image for a 7.8 nm HP MIL exposure on HSQ and compare the results with the calculated intensity. The simulated aerial image (Fig. 5**a**) is a grayscale depiction of the intensity computed using equation (8). Figure 5**b** shows an SEM image of HSQ lines with HP 7.8 nm patterned by MIL. The SEM image intensity (Fig. 5**c**) is a pixel-based calculation (grayscale value), computed as the average intensity along the lines.

The NILS number against the position of each line is plotted in Fig. 5**d** for a bandwidth of 4% that characterizes the beam used in our experiments. Here, one can see a practical demonstration of the variable NILS feature that was introduced in Section 3. For example, if the optical system of an industrial EUV lithography tool can project a 15.6 nm pitch aerial image with NILS 2.4, one has to study the 7[th] pair of lines that are positioned 105 nm around the centerline. As opposed to EUV-IL with transmission gratings, that usually patterns square areas measuring thousands of lines with NILS $\pi$, the only equivalent line in a MIL exposure is the centerline. However, the length of this line is a device-design parameter and can be as long as the diameter of the beam. In our case, the beam intensity is homogeneous within 1.5 mm, therefore, adequate statistical data (several SEM images) can be gathered along the length of the lines.

Finally, Fig. 5**e** shows the calculated contrast together with the SEM image contrast. Qualitatively, both curves exhibit a similar trend, with a decline in contrast as we move away from the centerline. The match is not perfect as the contrast obtained from the SEM image is a convolution of the aerial image, the resist contrast and the SEM electron beam profile.

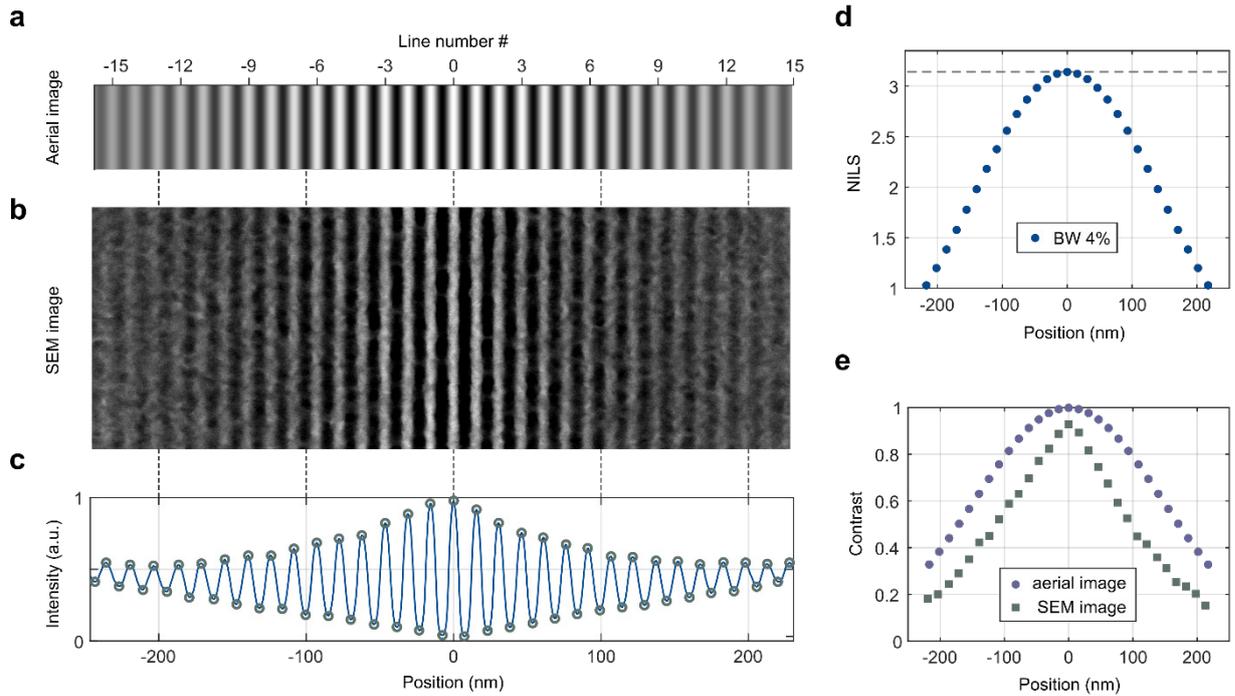

**Figure 5 | Analysis of a MIL exposure.** (**a**) Calculated aerial image for a HP 7.8 nm MIL device. (**b**) The corresponding SEM image of HSQ resist showing line/space patterning. The significantly expanded imaging area, compared to the ones in Fig. 4, includes lines with profoundly reduced contrast, as predicted by our calculated aerial image. (**c**) SEM image intensity, averaged along the direction of the lines, plotted versus the position from the centerline. (**d**) Calculated NILS numbers for an EUV beam with 4% bandwidth, as is the case for the beam used in our experiments, against the position of each line in the imaging field. (**e**) Contrast of the computed aerial image and the SEM image. The former is calculated from the simulated data shown in **a** and the latter from the intensity peaks in **c** that correspond to the SEM image.

**MIL with shorter wavelengths (Beyond EUV light)**

A notable advantage of utilizing a synchrotron source and an undulator as an insertion device is the flexibility one has in setting the wavelength shorter than the industrial standard of 13.5 nm. This could provide intriguing insights into the future of EUV lithography, where the adoption of shorter wavelengths might be considered as a means to enhance resolution. Although a wavelength reduction is not foreseeable in the near future, due to complexity in designing suitable optics, there is ongoing research towards this direction[55,56]. The XIL-II beamline features a tunable undulator that can be configured to any desired wavelength at the EUV and beyond EUV (BEUV) spectrum between 18 and 2.5 nm[50].

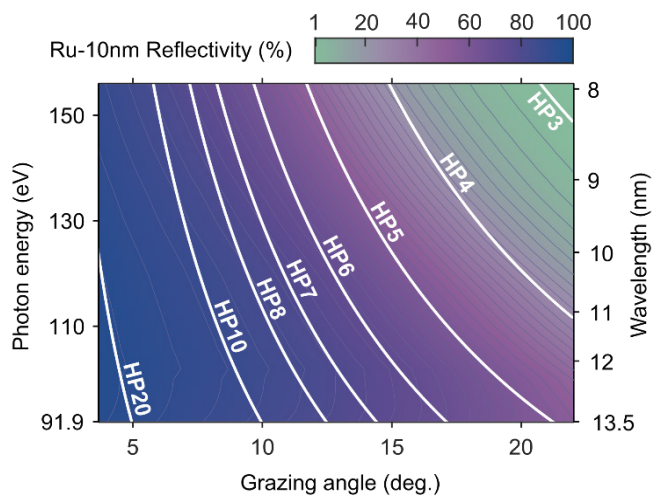

**Figure 6 | MIL at the BEUV spectrum.** Calculated reflectivity (%) for a 10 nm Ru film as a function of photon energy and grazing angle[35]. White lines show the combinations of grazing angles and wavelengths that produce the marked HP values.

According to equation (3), the pitch exhibits a linear correlation with the wavelength at a given angle. Figure 6 illustrates the calculated reflectivity for identical Ru mirrors as a function of both wavelength and grazing angle. The resulting HP, marked by white lines, represents a combination of grazing angle and wavelength. Consequently, this allows for the investigation of photoresist characteristics using light with shorter wavelengths, as well as the exploration of even higher resolutions. In fact, we further confine the areal image to 4 nm HP by utilizing photons with a wavelength of 10.8 nm, as shown in Fig. 7, along with other combinations of MIL devices and wavelengths. While the results appear to be of comparable quality to those obtained using the 13.5 nm, it becomes evident that achieving a HP of 4 nm exceeds the capabilities of either the resist, the applied process, or potentially both.

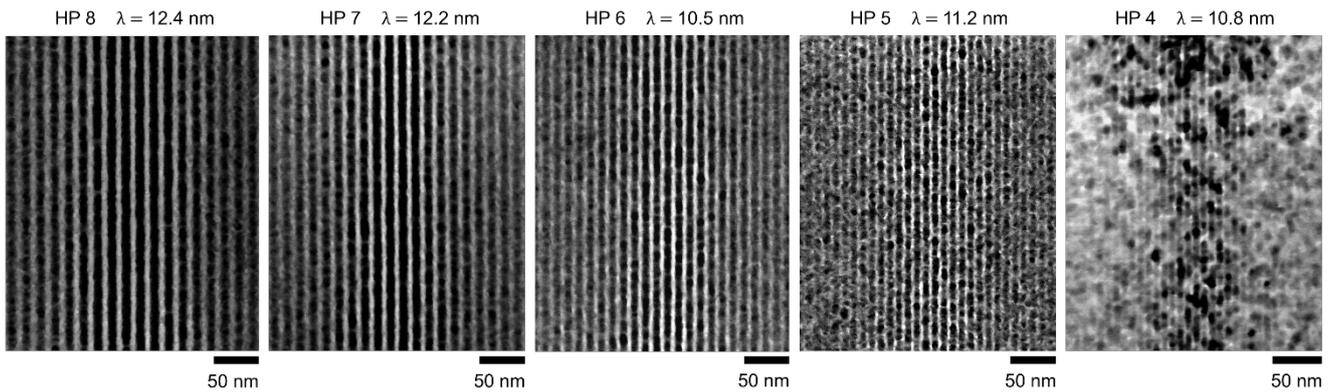

| HP 8   $\lambda = 12.4$ nm | HP 7   $\lambda = 12.2$ nm | HP 6   $\lambda = 10.5$ nm | HP 5   $\lambda = 11.2$ nm | HP 4   $\lambda = 10.8$ nm |

**Figure 7| MIL exposures at shorter wavelengths.** SEM images of HSQ photoresist lines of HP between 4 and 8 nm obtained with wavelengths below 13.5 nm. According to equation (3), HP depends on the grazing angle $\alpha$ and the wavelength $\lambda$. By tuning the undulator of the beamline, one can set the wavelength so that it gives the desired HP for a given MIL device.

### Conclusions and outlook

EUV lithography has become an integral part of chip fabrication with a profound and lasting impact on the continuous shrinkage roadmap. Recent developments in high-NA systems, expected to enter HVM in the next few years, along with ongoing research into hyper-NA systems, have opened new avenues for pushing the boundaries of lithographic resolution. In this challenging context, we have shown that there is a lot of room for further miniaturization until the ultimate photon-related resolution. Our EUV mirror interference lithography has demonstrated 5 nm HP resolution for photon-based lithography and paves the way for the development of systems with higher NAs or BEUV wavelengths towards ultimate resolution. The sub-10 nm HP capabilities and the on-demand selection of the tool-matching NILS number make it an ideal platform for scientific and industrial applications and notably for the development of photoresist and underlayer materials for future technology nodes. Crucially, the demonstration of 5 and 4 nm HP using 13.5 and 10.8 nm wavelengths, shows conclusively that photons themselves are not the bottleneck for resolution. Instead, our work shifts the focus towards developing new types of photoresist materials and optimizing the existing platforms.

Regarding the MIL device, the use of new mirror materials such as multilayers optimized for high grazing angles in EUV, holds great promise to significantly enhance the efficiency by boosting the reflectivity beyond that of these prototypes. This is important especially for high-resolution HPs and shorter wavelengths, where the Ru mirror reflectivity drops. An alternative MIL device design could leverage the highly advanced fields of micro electro-mechanical systems (MEMS) fabrication and wafer-level optics (WLO). This design could enable in-situ pitch tuning by adjusting the relative positions of the mirrors in response to configuration signals and enhance both throughput and flexibility. Finally, smaller bandwidths can be achieved with the implementation of an improved undulator and temporal filters. That would increase the number of lines with high contrast, a critical aspect for certain applications that require extensive area patterning. In conclusion, we hold a strong belief that MIL stands as a pivotal technology and a vital asset of EUV lithography, one that can drive the industry to greater resolutions and contribute to the leading semiconductor manufacturers in achieving the future technological milestones.

## Methods

### XIL-II beamline at the SLS

The XIL-II beamline is specifically designed for EUV and soft x-ray interference lithography and metrology at the SLS synchrotron facility. The electron beam in the storage ring measures 400 mA at 2.4 GeV energy. The beam exits the ring and goes through an undulator with 22 magnet pairs. The energy of the emitted coherent radiation can be adjusted between 70 and 500 eV by changing the undulator gap. Before reaching the endstation the light is reflected by 3 water-cooled mirrors with specific coatings and grazing angles, designed for distinct purposes such as alignment, focusing, and high-harmonics suppression[50]. The beam is focused on a user-adjustable pinhole that serves as a spatial filter (see Fig. 1**a**), while the whole ensemble operates at ultra-high vacuum conditions ($10^{-9}$ mbar). Finally, the light enters the endstation which is operated at a pressure of $10^{-7}$ mbar, where both the mask and the sample are positioned in a finely controlled distance. The photon flux is measured by calibrated photodiodes and the required dose is delivered by a high-speed mechanical shutter. The endstation is isolated by external vibrations, as it lies on an actively damped optical table.

### Device fabrication

A MIL device consists of 2 parts, namely the metallic frame and the mirrors. The former is a micromachined component designed in the Paul Scherrer Institute and crafted by AND Boumi AG in Switzerland using high precision manufacturing techniques. More specifically, the frame is a 10x10x4 mm unibody made by a solution-annealed Ti alloy (grade 5: Al 5%, V 4%). Shaped with a mixture of milling and wire-erosion which is a variation of electrical discharge machining (W-EDM), the minimum feature size is 0.06 mm (radius) and the minimum material thickness 0.1 mm at any point. The photonstop measures 0.25 mm and each of the two slits that form the beams are 0.125 mm wide.

**EUV mirror fabrication**

Replaceable EUV mirrors are fabricated on double-polished Si wafers of 250 $\mu$m thickness. The reflective surface is a 10 nm Ru thin film that is deposited by electron beam evaporation (Evatec BAK Uni). The deposition process was optimized for low surface roughness (root mean square: 0.2 nm) to prevent reflectivity losses.

**Photoresist**

The presented exposures are performed on HSQ films formed by diluting 6% XR1541 HSQ resist (DuPont) in methyl isobutyl ketone (MIBK, Technic France) at a ratio of 1:8. The thin resist films were prepared by spin-coating at 5000 rpm for 60 s and exposed without any post-application bake. Profilometry (Veeco DEKTAK 8) showed 13 nm thickness after development. Prior to coating, the Si wafers are treated with $O_2$ plasma for 2 min (TePla 300 plasma processor) set at 160 W RF power with a gas flow of 150 sccm. For the development of HSQ we used a commercial sodium hydroxide-based developer (AZ 351B, Merck) diluted 1:3 in deionized $H_2O$ for 35 s, a standard process for high resolution and high contrast patterning.

**SEM imaging**

The images in Fig. 4, 5 and 7 are obtained by a Hitachi Regulus 8230 ultra-high resolution SEM tool. The working distance is 1.7 mm, the magnification 250,000 times and the pixel size is 0.4 nm. The landing voltage was set at 0.5 kV using a high deceleration of 3.5 kV, with the probe current kept at 10 pA. The image is acquired with a charge-suppressed scan and the collected signal is a mix of the secondary and the backscattered electrons.

**Acknowledgements**

The authors would like to thank Renzo Rotundo and Markus Kropf for their technical support.

**Author contributions**

Y.E., I.M. and D.K. conceived the MIL device and its working principles. M.V. supervised the fabrication and the assembly of the devices. I.G. performed the experiments, the SEM imaging and analyzed the data. I.G., D.K. and I.M. performed the numerical simulations. I.G. wrote the manuscript with inputs from all the authors. D.K. and Y.E. supervised the work.

**Additional information**

**Supplementary information** accompanies this paper at _______

**Competing financial interests**: The authors declare no competing financial interests.

**Supplementary Information**

## A. EUV-IL based on transmission gratings

In EUV-IL, an interference pattern created by two or more coherent beams is used to expose a photoresist. A synchrotron beamline (Fig. S1**a**) is an ideal light source for such systems. We use masks with transmission diffraction gratings that are fabricated on thin transparent membranes (Fig. S1**b**). The standing wave that results from the interference of two or more mutually coherent beams, generates a sinusoidal intensity variation that exposes the sample photoresist. In the simplest case of a two-beam interference (Fig. S1**c**), the light beam with wavelength $\lambda$ passes through a mask that consists of two gratings with period $g$ surrounded by a photonstop metal layer. Each order of diffraction $m$ is deflected at a specific angle $a_m$ with respect to the incident, given by the grating equation:

$$\sin a_m = \frac{m\lambda}{g} \tag{S1}$$

The interference period depends on the angle of the diffracted beams $\theta_m$ and is given by Eq. (S2). However, the latter is independent of the wavelength, because $\theta_m = 2\alpha_m$ and therefore Eq. (S1) & (S2) are combined in Eq. (S3).

$$P = \frac{\lambda/2}{\sin(\theta_m/2)} \tag{S2}$$

$$P = \frac{\lambda/2}{\sin a_m} = \frac{\lambda/2}{m\lambda/g} = \frac{g}{2m} \tag{S3}$$

Here, one can see the frequency multiplication, one of the main advantages of interference lithography. Our system utilizes the 1st order diffracted beams, because the diffraction efficiency of higher orders reduces dramatically, e.g. the 2nd order has almost 100x lower efficiency[1]. In 1st order diffraction shown in Fig. S1**c**, the patterning pitch is half of the one required to fabricate the used mask.

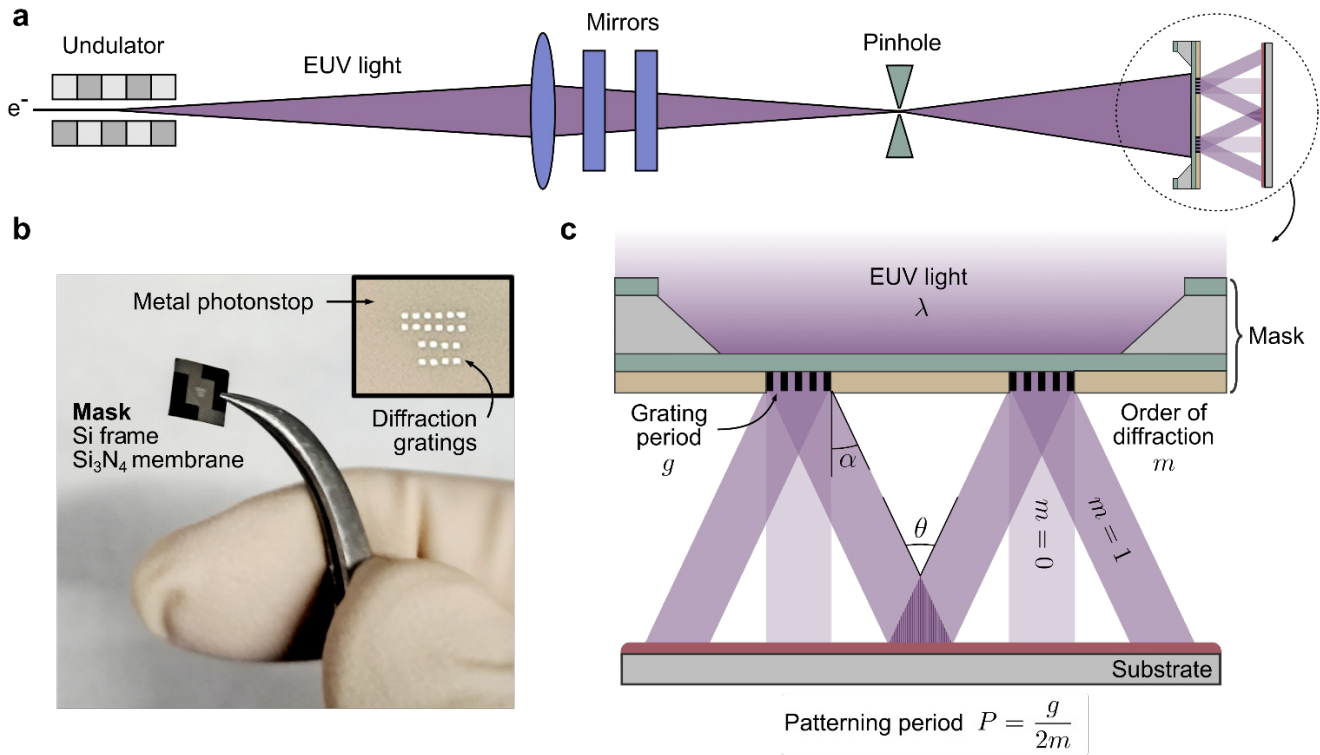

**Figure S1 | EUV interference lithography.** (**a**) Schematic representation of the XIL-II beamline at the Paul Scherrer Institute. EUV light with tunable wavelength is generated by an undulator and gets reflected off a series of mirrors for high harmonic suppression and focusing. The beam is then focused on a pinhole (spatial filter) and the spatially coherent beam subsequently illuminates the imaging module (mask with gratings). (**b**) Photograph of a mask with transmission gratings. The spacing between the grating sets varies with their grating period in order to get the same mask-sample spacing for the different diffraction angles. (**c**) The operation principle of EUV-IL based on transmission gratings. A silicon nitride membrane (green) is grown and released by KOH-etching of the Si wafer (gray). Gratings with period $g$ are formed on HSQ photoresist by EBL. The metal photonstop (beige) ensures that light passes only through the gratings where it gets diffracted. The interference pattern of the 1$^{st}$ order diffracted beams gets recorded on the sample photoresist (red) and its period is independent of the wavelength $\lambda$.

**B. The period of a MIL exposure**

The interference pattern intensity as a function of the mirror angle and the wavelength is given by Eq. (2)[2]. The pitch of a MIL exposure in Eq. (4),(S5) is derived by Eq. (2) using Euler's formula and trigonometric identities $\cos(\sin(-u)) = \cos(\sin(u))$ & $\sin(\sin(-u)) = -\sin(\sin(u))$, as following:

$$I(x) = A^2 \left| e^{i\frac{2\pi x}{\lambda}sin(2a)} + e^{i\frac{2\pi x}{\lambda}sin(-2a)} \right|^2$$

$$I(x) = 4A^2 \cos^2\left(\frac{2\pi x}{\lambda}sin(2a)\right)$$

$$I(x) = 2A^2 \left(1 + cos\left(\frac{4\pi}{\lambda}sin(2a)\,x\right)\right). \tag{S4}$$

The frequency of the periodic Eq. (S4) is $f = (4\pi/\lambda)\sin 2\alpha$, therefore, the period $P = 2\pi/f$ is:

$$P = \frac{\lambda}{2\sin 2\alpha} \tag{S5}$$

The distance between the centers of the reflection and the overlap areas (Fig. **1b** & S2**a**) is:

$$S = \frac{d}{2\tan 2\alpha} \tag{S6}$$

**C. MIL with a non-ideal setup**

In the following paragraphs we extend the concept of MIL and we study its behavior in the non-ideal scenario where there is an asymmetry in the grazing angles of the two mirrors and a possible wafer misalignment. In the special case shown in Fig. **1b**, one could introduce an additional angle component $\delta$ on the second mirror. Additionally, the normal vector of the wafer plane could also form an angle $\theta$ with the direction of propagation of the incoming beam $z$ (Fig. S2**a**). The case of $\delta \neq 0$ is very probable and can vary every time new mirrors are glued on the device. To this end, we derive Eq. (S12) that shows the corrected pitch based on the geometrical analysis of Fig. S2**b**.

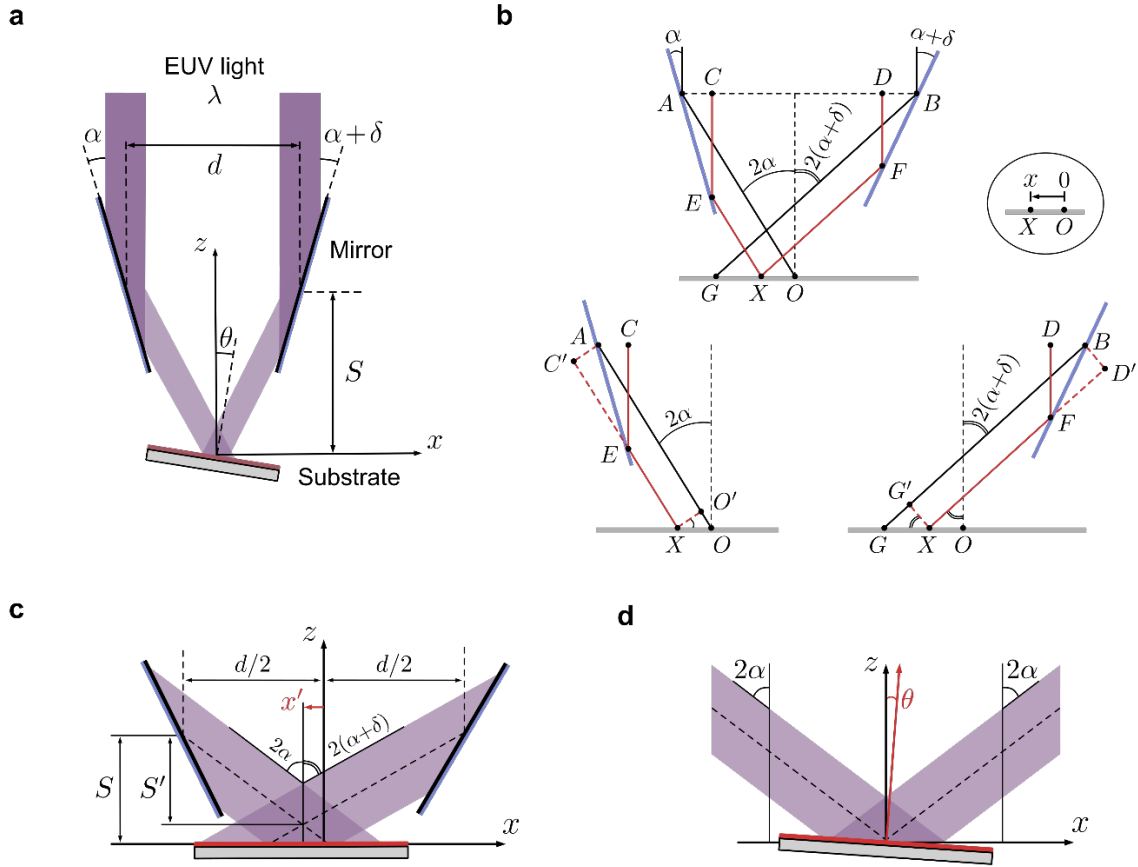

**Figure S2 | Geometrical analysis of MIL.** (**a**) Extended version of Fig. 1**b** featuring a non-symmetrical case of the MIL device. One of the mirrors has a higher angle than the other by a value $\delta$, and the wafer is tilted at an angle $\theta$. (**b**) Geometrical analysis of the asymmetric case, in which the difference of the optical path lengths $CEX$ and $DFX$ is expressed as a function of position $x$ and the grazing angles in order to derive the interference pitch. (**c**) The elevation of the overlap area with respect the sample at the asymmetric case. (**d**) The tilted-wafer case, in which the pitch is projected at an angle $\theta$ leading to an observed pitch expanded by $1/\cos\theta$.

We define as positive $\delta$ the angle that increases the value of $\alpha$. To understand the effect on the patterning pitch, one can express the difference of the optical path lengths of the two beams that interfere at point $X$ of Fig. 2**b**, $CEX$ and $DFX$ (red lines). We assume that the wavefront reaches the two mirrors at points A and B respectively. The light reflected at point $A$ reaches the wafer plane at point $O$ that we assume as the origin ($x = 0$) and the distance $OX$ can be parametrized as $x$. We express the optical path lengths $OPD1 = AO - CEX$ and $OPD2 = BG - DFX$ and study them individually in the assistive schematics of Fig. 2**b**, where we substitute $C$ and $D$ with the virtual sources $C'$ and $D'$, positioned so that $CE = C'E$ and $DE = D'E$. We now observe that $OPD1 = AO - C'X = OO'$ and $OPD2 = BG - D'X = GG'$ that can be expressed as functions of $x$ and the grazing angles:

$$OO' = x \sin 2\alpha \qquad (S7)$$

$$GG' = (OG - x) \sin 2(\alpha + \delta) \qquad (S8)$$

The optical path difference between the red lines $DFX$ and $CEX$ is:

$$OPD = DFX - CEX = (BG - OPD2) - (AO - OPD1) = (BG - GG') - (AO - OO')$$

Using Eq. (S7) and (S8)

$$OPD = BG - OG \sin 2(\alpha + \delta) + x \sin 2(\alpha + \delta) - AO + x \sin 2\alpha$$

$$OPD = (\sin 2\alpha + \sin 2(\alpha + \delta))x + (BG - AO - OG \sin 2(\alpha + \delta)) \qquad (S9)$$

Constructive interference occurs when $OPD = n\lambda$. While constant factor $(BG - AO - OG \sin 2(\alpha + \delta))$ in Eq. (S9) can be neglected because it only offsets the interference pattern, the multiplier of $x$ defines the interference period (pitch) as:

$$P = \frac{\lambda}{(\sin 2\alpha + \sin 2(\alpha + \delta))} \qquad (S10)$$

The trigonometric formula $\sin u + \sin v = 2 \sin \frac{u+v}{2} \cos \frac{u-v}{2}$ reduce the denominator of Eq. (S10) to:

$$\sin 2\alpha + \sin 2(\alpha + \delta) = 2 \sin(2\alpha + \delta) \cos \delta \qquad (S11)$$

Combining Eq. (S10) & (S11) one gets the period of MIL for the asymmetric case in Eq. (S12) that reduces to Eq. (S5) for $\delta \approx 0$. This assumption corresponds to any practical misalignment in our system and justifies the use of Eq. (S5).

$$P = \frac{\lambda}{2 \sin(2a + \delta) \cos \delta} \qquad (S12)$$

The distance $S$ between the reflection point and the interference area is changing as well. Interference happens along the rhomboidal area of beam overlap, which is the reason behind the depth-of-focus tolerance of EUV-IL. However, the ideal device-wafer distance is the one that aligns the overlap center with the sample. We study the case of positive $\delta$ in Fig. 2c where the center is shifted by $x'$ and its new distance is $S'$. These parameters can be expressed as functions of the grazing angles and the mirror spacing in Eq. (S13) and (S14). Adding those and solving for $S'$ gives the corrected distance in Eq. (S15), which reduces to Eq. (S6) for $\delta \approx 0$.

$$\tan\big(2(\alpha + \delta)\big) = \frac{d/2 + x'}{S'} \tag{S13}$$

$$\tan 2\alpha = \frac{d/2 - x'}{S'} \tag{S14}$$

$$S' = \frac{d}{\tan 2\alpha + \tan\big(2(\alpha + \delta)\big)} \tag{S15}$$

A wafer tilt is expected to have a negligible effect even in an experimental setup. We introduce an angle of $\theta$ with respect to the ideal wafer normal, as shown in Fig. S2**d**. According to Eq. (S16) the observed pitch is increased by the projection factor $1/\cos\theta$.

$$P_\theta = \frac{P}{\cos\theta} \tag{S16}$$

## D. NILS calculation for monochromatic light

Normalized image log-slope (NILS) characterizes the edge definition of a pattern. It is defined in Eq. (S17)[3] as the intensity slope at the border of the pattern area, normalized by the intensity value at this point and the nominal linewidth of the patterned feature.

$$\mathrm{NILS} = \frac{1}{I}\frac{dI}{dx}w \tag{S17}$$

Plugging Eq. (S4) in Eq. (S17) yields:

$$\frac{1}{I}\frac{dI}{dx}w = -\frac{4\pi}{\lambda}sin(2a)\frac{sin\left(\frac{4\pi}{\lambda}sin(2a)\,x\right)}{1 + cos\left(\frac{4\pi}{\lambda}sin(2a)\,x\right)}w$$

Using the trigonometric identity: $\quad \dfrac{\sin\theta}{1 + \cos\theta} = \tan\dfrac{\theta}{2}$

$$\frac{1}{I}\frac{dI}{dx}w = -\frac{4\pi}{\lambda}sin(2a)\,tan\left(\frac{2\pi}{\lambda}sin(2a)\,x\right)w \tag{S18}$$

We calculate NILS for HP lines with width:

$$w = \frac{p}{2} = \frac{\lambda}{4\sin(2a)} \qquad (S19)$$

Eq. (S19) in Eq. (S18) gives:

$$\frac{1}{I}\frac{dI}{dx}w = -\frac{\lambda}{4\sin(2a)}\frac{4\pi}{\lambda}sin(2a)\,tan\left(\frac{2\pi}{\lambda}sin(2a)\,x\right) = -\pi\,tan\left(\frac{2\pi}{\lambda}sin(2a)\,x\right)$$

Including the pitch from Eq. (S5), NILS expression becomes:

$$\text{NILS} = \frac{1}{I}\frac{dI}{dx}w = -\pi\,tan\left(\frac{\pi}{p}x\right) \qquad (S20)$$

We calculate NILS specifically at the borders of the HP lines. Therefore, the positions $x$ are given by the sequence in Eq. (S21) that becomes Eq. (S22) including the pitch Eq. (S5).

$$x = (2n-1)\frac{P}{4}, \qquad n = 1,2,\dots \qquad (S21)$$

$$x = (2n-1)\frac{\lambda}{8\sin(2a)}, \qquad n = 1,2,\dots \qquad (S22)$$

Finally, Eq. (S22) in Eq. (S20) gives:

$$\text{NILS} = \frac{1}{I}\frac{dI}{dx}w = -\pi\,tan\left(\frac{2\pi}{\lambda}sin(2a)\,(2n-1)\frac{\lambda}{8\sin(2a)}\right) = -\pi\tan\left(\frac{\pi}{4}(2n-1)\right), \quad n = 1,2,\dots$$

Signs distinguish the leading from the trailing edges, but NILS is reported as an absolute value, Eq. (S23).

$$\text{NILS} = |(-1)^n\,\pi|, \quad n = 1,2,\dots \;\rightarrow\; \text{NILS} = \pi \qquad (S23)$$